\documentclass[prl,twocolumn,amssymb,showpacs,superscriptaddress]{revtex4}
\usepackage{bm}
\usepackage{graphicx}
\begin{document}
\title{Local density of states subject to finite impurity
       concentration in graphene}
\author{Yuriy~V.~Skrypnyk}
\affiliation{Kurdyumov Institute of Metal Physics,
             National Academy of Sciences of Ukraine,
             Vernadskogo 36, Kyiv 03142, Ukraine}
\author{Vadim~M.~Loktev}
\affiliation{Bogolyubov Institute for Theoretical Physics,
             National Academy of Sciences of Ukraine,
             Metrolohichna 14-b, Kyiv 03143, Ukraine}

\begin{abstract}
It is demonstrated that there is a characteristic impurity
concentration, at which variation with concentration and overall
appearance of the local density of states at the impurity site in
graphene are changing their behavior. Features that are prominent in
the local density of states for the single impurity are disappearing
from it when impurity concentration far exceeds this critical
value. The impurity subsystem not only induces the rearrangement of
the electron spectrum in graphene, but also undergoes a substantial
spectral transformation by itself, which can be observed
experimentally.
\end{abstract}


\maketitle

\paragraph{Introduction.}
The spectrum rearrangement, which is inherent in disordered systems,
have been studied for a large variety of systems over a period of last
decades \cite{ilp,lif,kos}. In essence, this phenomenon is grounded on
the indirect interaction of impurities through the host system. The
impurity state that is formed close to the band edge, or in the
vicinity of any other van Hove singularity, has an effective radius
that is large compared to the lattice constant. Thus, the spatial
overlap of such impurity states occurs at the low impurity
concentration ($c\ll 1$). This overlap results in a radical alteration
of spectral properties of the disordered system, which is especially
pronounced near the energy of the impurity state.

Predominantly, the spectrum rearrangement is talked about with regard
to a remarkable change in the dispersion relation or in the total
density of states. However, with recent advances in the scanning
tunneling spectroscopy (STS), it became possible to observe directly
the local density of states (LDOS) \cite{bal,nak,fang}. Thus, it seems
reasonable to inquire how the spectrum rearrangement might be
reflected in the shape of the LDOS.

Below we would try to examine whether an increase in the impurity
concentration influence the behavior of the LDOS to such an extent, so
that it is permissible to speak about the impurity induced
rearrangement of the LDOS on its own account. It is well known that
an increase in the impurity concentration throws a veil over the LDOS
features that correspond to a single impurity in the system
\cite{v1,v2,v3}. In this connection, it should be stressed that we
will be well within our rights introducing the concept of the impurity
induced rearrangement of the LDOS only in a case, when qualitatively
different regimes in the LDOS behavior can be distinguished, and the
critical concentration that indicates the transition from one regime
to another is closely related to the spatial overlap of the impurity
states.

Our treatment of this issue is based on a straightforward model for
graphene \cite{gr2}. Because graphene is a purely two--dimensional
(2D) object, the STS measurements of the electronic LDOS in this
material are decidedly natural. Moreover, possibilities of the
resonance state formation and respective features in the single
impurity LDOS at the impurity site and in its neighborhood have been
discussed elsewhere \cite{kivel,vitor,skrloc}. Opportunities to check
up by future STS experiments on these characteristics in the near
field of the impurity have been already addressed \cite{balcond}, and
first reports on the STS measurements in graphene have been published
recently \cite{big}.

\paragraph{Model.}
For the sake of simplicity we consider a substitutional binary alloy
with a diagonal disorder in the tight--binding approximation. This
uncomplicated model of a disordered system is attributed by tradition
to Lifshitz \cite{lm}. It features absolute randomness in space
distribution of impurities. We choose the asymmetric definition of
impurity perturbation. Consequently, the on--site potentials are
$V_{L}$ with the probability $c$, or $0$ otherwise. The corresponding
Hamiltonian reads,
\begin{equation}
\bm{H} = \bm{H}_0 +\bm{H}_{imp},\qquad%
\bm{H}_{imp}=V_{L}{\sum_{\bm{n},\alpha}}'%
c^{\dag}_{\bm{n}\alpha}c^{}_{\bm{n}\alpha},
\label{fullH}
\end{equation}
where $\bm{n}$ refers to lattice cells, $\alpha$ enumerates sublattices,
$c^{\dag}_{\bm{n}\alpha}$ and $c^{}_{\bm{n}\alpha}$ are electron
creation and annihilation operators. The summation in
Eq.~(\ref{fullH}) is restricted to those sites that are occupied by
impurities. Since the impurity perturbation is local in its character,
the main physics of graphene in the linear dispersion domain can be
captured by a model 2D system with a single Dirac cone in the
spectrum. Thus, the host Hamiltonian $\bm{H}_{0}$ can be
written in the following way, 
\begin{eqnarray}
\bm{H}_{0}&=&\sum_{\bm{k}}[f(\bm{k}) c^{\dag}_{1}(\bm{k})%
c^{}_{2}(\bm{k})+f^{*}(\bm{k}) c^{\dag}_{2}(\bm{k})
c^{}_{1}(\bm{k})], \label{h0} \\
c^{}_{\alpha}(\bm{k})&=&\frac{1}{\sqrt{N}}\sum_{\bm{n}}e^{i\bm{k}\bm{n}}%
c^{}_{\bm{n}\alpha},\quad f(\bm{k})=\frac{a}{2\sqrt{\pi}}(k_{x}+i k_{y}),%
\nonumber
\end{eqnarray}
where the magnitude of the hopping parameter is chosen so that the
bandwidth is unity, when the Brillouin zone is approximated with a
circle. Then, the diagonal element of the host Green's function (GF)
$\bm{g}=(\epsilon-\bm{H}_{0})^{-1}$ in the vicinity of the Dirac
point is given by \cite{skrloc,per}  
\begin{equation}
g_{\bm{n}\alpha\bm{n}\alpha}(\epsilon)\equiv g_{0}(\epsilon)\approx%
2\epsilon\ln\left|\epsilon\right|-i\pi\left|\epsilon\right|,\quad%
\left|\epsilon\right|\ll 1. \label{gf}
\end{equation}

\paragraph{Single impurity problem.}
In a case, when only a single impurity is present in the lattice, the
diagonal element of the GF $\bm{\mathcal{G}}=(\epsilon-\bm H)^{-1}$ at
the impurity site becomes 
\begin{equation}
\mathcal{G}_{0}(\epsilon)=g_{0}(\epsilon)/[1-V_{L}g_{0}(\epsilon)],
\label{sigf}
\end{equation}
For a sufficiently large impurity perturbation $V_{L}$, the respective
single impurity local density of states (LDOS) $\rho_{imp}(\epsilon)=%
-\pi^{-1}\mathop{\mathrm{Im}}\mathcal{G}_{0}(\epsilon)$ manifests a
resonance peak \cite{skrloc}. This peak is analogous in its nature to
the resonances described in unconventional superconductors
\cite{bal}. Its energy  is provided by the Lifshitz equation,
\begin{equation}
1=V_{L}\mathop{\mathrm{Re}}g_{0}(\epsilon_{r})%
\approx 2V_{L}\epsilon_{r}\ln\left|\epsilon_{r}\right|. \label{le}
\end{equation}
The solution of this equation, i.e., the resonance energy
$\epsilon_{r}$ is located above the Dirac point in the spectrum for
perturbation $V_{L}<0$, and vice versa. This property of the resonance
state holds valid for any two symmetric bands touching each other at a
certain energy. By grouping Eqs.~(\ref{gf}) and (\ref{sigf}) together,
obtain,
\begin{equation}
\rho_{imp}(\epsilon)=|\epsilon|/[%
(1-2V_{L}\epsilon\ln|\epsilon|)^{2}+(\pi V_{L}\epsilon)^{2}].
\label{sinldos}
\end{equation}
Power series expansion of the denominator in Eq.~(\ref{sinldos})
about the resonance energy yields,  
\begin{eqnarray}
\rho_{imp}(\epsilon)&\approx&|\epsilon|\Gamma_{r}^{2}/%
\{[\pi V_{L}\epsilon_{r}]^{2}\left[(\epsilon-\epsilon_{r})^{2}+%
\Gamma_{r}^{2}\right]\},\nonumber \\
\Gamma_{r}&=&\pi|\epsilon_{r}|/\left[2|1+\ln|\epsilon_{r}||\right],
\label{gam}
\end{eqnarray}
Thus, the LDOS has the Lorentz profile in the vicinity of the
resonance, when it is located so close to the Dirac point, that the
inequality
\begin{equation}
\gamma_{r}\equiv\frac{\Gamma_{r}}{|\epsilon_{r}|}\approx\frac{\pi}%
{2\left|1+\ln|\epsilon_{r}|\right|}\ll 1 \label{wd}
\end{equation} 
is fulfilled. As a rule, a resonance is accepted as a well--defined,
when this condition is met.

\begin{figure}
\includegraphics[width=0.475\textwidth,clip]{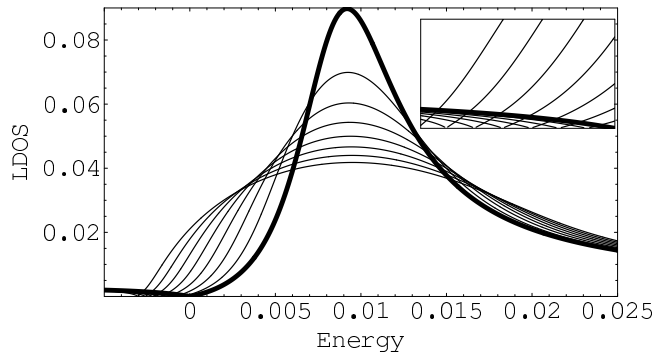}
\caption{\label{f1} The LDOS at the impurity site for
  $\epsilon_r=0.01$ at $c=4n\times10^{-5},\, n=1,2\dots7$. The single
  impurity LDOS is displayed for comparison by the thick line. The
  peak height is decreasing with increasing the concentration. The
  inset is showing the bottom left hand corner at a larger scale.}
\end{figure}

\begin{figure}
\includegraphics[width=0.475\textwidth,clip]{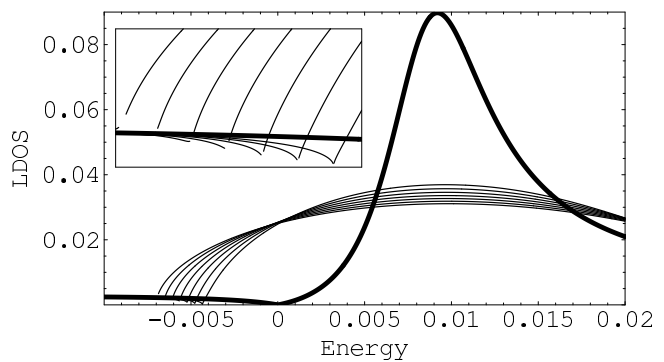}
\caption{\label{f2} The LDOS at the impurity site for
  $\epsilon_r=0.01$ at $c=(36+4n)\times10^{-5},\, n=1,2\dots7$. The single
  impurity LDOS is displayed for comparison by the thick line. The
  peak height is decreasing with increasing the concentration. The
  inset is showing the bottom left hand corner at a larger scale.}
\end{figure}

\paragraph{Finite impurity concentration.}
It is convenient to consider conditional (or weighted) GFs when
dealing with the finite impurity concentration. The conditional GF
with the first site occupied by an impurity,
\begin{equation}
\bm{\mathcal{G}}^{(imp,host)}=V_{L}^{-1}\bm{H}_{imp}\bm{\mathcal{G}},
\end{equation}
can be represented through the self--energy after averaging over
possible impurity distributions \cite{ellkr,ell},
\begin{equation}
\bm{G}^{(imp,host)}=V_{L}^{-1}\bm{\Sigma}\bm{G},\qquad%
\bm{G}=\bm{g}+\bm{g}\bm{\Sigma}\bm{G},
\end{equation}
where $\bm{G}=\biglb<\bm{\mathcal{G}}\bigrb>$ is the averaged GF of
the system. As long as the impurity concentration remains sufficiently
small to neglect multiple occupancy corrections, the self--energy can
be approximated by means of the well--known modified propagator method
\cite{langer},
\begin{equation}
\bm{\Sigma}\approx\sigma\bm{I},\qquad%
\sigma=cV_{L}/[1-V_{L}g_{0}(\epsilon-\sigma)].\label{mp}
\end{equation}
The self--energy, which is identical on both sublattices owing to the
symmetry, is also site--diagonal within this approximation. In order
to obtain a quantity that can be related to $\mathcal{G}_0(\epsilon)$
in a single impurity problem, the diagonal in lattice indices element
of the conditional GF should be properly scaled with the impurity
concentration,
\begin{equation}
c^{-1}G^{(imp,host)}_{0}\approx g_{0}(\epsilon-\sigma)/%
[1-V_{L}g_{0}(\epsilon-\sigma)].
\end{equation}
Thus, we get an expression that formally resembles Eq.~(\ref{sigf}),
in which the host GF $\bm{g}$ is replaced by the propagator of the
disordered system $\bm{G}(\epsilon)\approx\bm{g}(\epsilon-\sigma)$.
Finally, the average LDOS at the impurity site can be written down
through the self--energy,
\begin{eqnarray}
&\rho_{loc}(\epsilon)&\approx -\frac{1}{\pi}\mathop{\mathrm{Im}}%
\bigl[\frac{g_{0}(\epsilon-\sigma)}{1-V_{L}g_{0}(\epsilon-\sigma)}%
+\frac{1}{V_{L}}\bigr]= \nonumber \\
&=&\!\!\!\!\!\!\! -\frac{1}{\pi V_{L}}\mathop{\mathrm{Im}}\frac{1}%
{1-V_{L}g_{0}(\epsilon-\sigma)}=-\frac{1}{\pi}%
\frac{\mathop{\mathrm{Im}}\sigma}{c V_{L}^{2}}. \label{ldos}
\end{eqnarray}     
With the help of the standard substitution,
\begin{equation}
\epsilon-\sigma=\varkappa \exp(i\varphi),\quad \varkappa>0,\quad%
0<\varphi<\pi, \label{subs}
\end{equation}
the imaginary part of the self--consistency condition in
Eq.~(\ref{mp}) becomes,
\begin{eqnarray}
&&cV_{L}^{2}\left[2\ln\varkappa+(2\varphi-\pi)\cot\varphi\right]+%
\nonumber\\
&+&\left[1-V_{L}\varkappa(2\ln\varkappa\cos\varphi-(2\varphi-\pi)%
\sin\varphi\right)]^{2}+ \nonumber \\
&+&\left[V_{L}\varkappa(2\ln\varkappa\sin\varphi+%
(2\varphi-\pi)\cos\varphi\right)]^{2}=0, \label{sim} 
\end{eqnarray}
where the particular form of the host GF (see Eq.~(\ref{gf})) has been
taken into account. At the given impurity perturbation $V_{L}$ and the
given impurity concentration $c$, this equation always has two
solutions for the phase $\varphi$, when $\varkappa$ exceeds some
threshold value. In a turn, both roots provide corresponding
magnitudes of the energy, when substituted into the real part of
Eq.~(\ref{mp}),
\begin{eqnarray}
\epsilon&=&\varkappa\cos\varphi+cV_{L}%
\left[1-V_{L}\varkappa(2\ln\varkappa\cos\varphi-\right. \nonumber \\
&-&\left.(2\varphi-\pi)\sin\varphi)\right]/\bigl\{%
\left[1-V_{L}\varkappa(2\ln\varkappa\cos\varphi\right.-\bigr. \nonumber \\
&-&\left.(2\varphi-\pi)\sin\varphi)\right]^{2}+%
\left[V_{L}\varkappa(2\ln\varkappa\sin\varphi\right. \nonumber \\
\bigl.&+&\left.(2\varphi-\pi)\cos\varphi)\right]^{2}\bigr\}.\label{sim2} 
\end{eqnarray}
The LDOS from Eq.~(\ref{ldos}) can be expressed in the same variables
too,
\begin{equation}
\rho_{loc}(\epsilon)=\varkappa\sin\varphi/(\pi c V_{L}^{2}).
\label{sim3}
\end{equation}
Several examples of the LDOS at the impurity site, which were
calculated according to Eqs.~(\ref{sim})-(\ref{sim3}), are depicted in
Figs.~\ref{f1}--\ref{f2} for a range of impurity concentrations.

\paragraph{Rearrangement of resonance peak.}
In a parallel to the single impurity problem, the resonance energy
that corresponds to the given impurity concentration should be a
solution of the equation (cp. to Eq.~(\ref{le})),
\begin{equation}
1=V_{L}\mathop{\mathrm{Re}}G_{0}(\epsilon_{r}(c)).
\end{equation}
By using Eq.~(\ref{gf}) and the substitution (\ref{subs}), this
equation can be recast as follows, 
\begin{eqnarray}
1&=&2V_{L}\epsilon_{r}(c)[\ln|\epsilon_{r}(c)/%
\cos\varphi_{r}(c)|+ \nonumber \\
&+&(\pi/2-\varphi_{r}(c))\tan\varphi_{r}(c)]. \label{resc}
\end{eqnarray}
From here on, we will assume without any loss of generality that
$\epsilon_{r}>0$ to make expressions more readable. When the
concentration is low enough to keep $\varphi_{r}(c)$ small, an
approximate solution of Eq.~(\ref{resc}) is: 
\begin{equation}
\epsilon_{r}(c)\approx\epsilon_{r}[1+\gamma_{r}\varphi_{r}(c)],%
\qquad \varphi_{r}(c)\ll 1. \label{offset}
\end{equation} 
Thus, because $\gamma_{r}$ (see Eq.~(\ref{wd})) is the resonance width
parameter that should be small for the well--defined quasilocalized state,
the energy of resonance varies slowly with $\varphi_{r}(c)$. The
concentration dependence of $\varphi_{r}(c)$ can be obtained from the
self--consistency condition (\ref{sim}). The second term in
Eq.~(\ref{sim}) is zero by the definition of $\epsilon_{r}(c)$, the
remaining two lead to the relation,
\begin{eqnarray}
c&=&-2\epsilon_{r}^{2}(c)\tan\varphi_{r}(c)\times  \label{cself}\\
&\times&(\ln|\epsilon_{r}(c)/\cos\varphi_{r}(c)|\tan\varphi_{r}(c)+%
\varphi_{r}(c)-\pi/2). \nonumber
\end{eqnarray} 
As above, Eq.~(\ref{cself}) considerably simplifies while the phase
at the resonance energy $\varphi_{r}(c)$ remains small,
\begin{equation}
c\approx\pi\epsilon_{r}^{2}(c)\varphi_{r}(c)%
(1+\varphi_{r}(c)/\gamma_{r}),\qquad \varphi_{r}(c)\ll 1.
\label{cselfsim}
\end{equation}
This relation reveals a hidden physical meaning of parameter
$\gamma_{r}$. It clearly shows that $\gamma_{r}$ is the characteristic
phase for the resonance state. The corresponding impurity
concentration is given by Eq.~(\ref{cselfsim}) with
$\varphi_{r}(c)=\gamma_{r}$,
\begin{equation}
c_{r}=-\frac{\pi^{2}\epsilon_{r}^{2}(c_{r})}{1+\ln\epsilon_{r}}%
\approx-\frac{\pi^{2}\epsilon_{r}^{2}}{1+\ln\epsilon_{r}}%
\approx-\frac{\pi^{2}\epsilon_{r}^{2}}{\ln\epsilon_{r}}.\label{cr}
\end{equation}
At the critical impurity concentration $c_{r}$ the damping of the
resonance induced by the disorder becomes equal to the damping of the
single impurity state, $-\mathop{\mathrm{Im}}\sigma\approx%
\epsilon_{r}(c_{r})\varphi_{r}(c_{r})\approx\Gamma_{r}$, and,
respectively, the LDOS at the resonance energy decreases to the one
half of its magnitude for the single impurity problem (see
Eq.~(\ref{gam})), 
\begin{equation}
\rho_{loc}(\epsilon_{r}(c_{r}))\approx%
\frac{\epsilon_{r}(c_{r})\varphi_{r}(c_{r})}{\pi c_{r}V_{L}^{2}}=%
\frac{1}{2}\frac{1}{\pi^{2}V_{L}^{2}\epsilon_{r}}%
=\frac{\rho_{imp}(\epsilon_{r})}{2}.\label{half}
\end{equation}

Neglecting the concentration offset in the resonance position
(\ref{offset}), Eq.~(\ref{cselfsim}) can be easily solved for the
phase,
\begin{equation}
\varphi_{r}(c)\approx(\gamma_{r}/2)(\sqrt{1+(8c/c_{r})}-1), \label{phase}
\end{equation}
Consequently, the concentration dependence of the resonance peak
height in the LDOS at the impurity site immediately follows from
Eqs.~(\ref{half}) and (\ref{phase}),
\begin{equation}
\rho_{loc}(\epsilon_{r}(c))\approx\frac{\sqrt{1+(8c/c_{r})}-1}%
{(4c/c_{r})}\rho_{imp}(\epsilon_{r}).
\end{equation}
With increasing impurity concentration, this height decreases linearly
with concentration at the start,
\begin{equation}
\rho_{loc}(\epsilon_{r}(c))\approx[1-(2c/c_{r})]%
\rho_{imp}(\epsilon_{r}),\qquad c\ll c_{r},
\end{equation}
but then the rate of this gradual decrease slows down,
\begin{equation}
\rho_{loc}(\epsilon_{r}(c))\approx\sqrt{c_{r}/(2c)}%
\rho_{imp}(\epsilon_{r}),\qquad c\gg c_{r}.
\end{equation}
The difference between these two regimes is clearly evident from the
comparison of Fig.~\ref{f1} with Fig.~\ref{f2}. In principle, it
should be feasible to capture  the change of the rate at which the
magnitude of the LDOS at the resonance energy varies with the impurity
concentration by STS measurements, not to mention the halving of the
resonance peak height. Thus, we expect that the task of pinpointing
the critical concentration of the spectrum rearrangement, which is a
valuable parameter of the system, is accessible by actual experiments.

\paragraph{Rearrangement of antiresonance.}
Apart from the resonance, there is also a noticeable dip in the LDOS
shape close to the Dirac point in the host system, which can be
considered like a kind of an antiresonance. At low impurity
concentration the position of this dip practically coincides with
those energy in the spectrum, at which $\varphi=\pi/2$. At this energy
the imaginary part of the self--consistency condition Eq.~(\ref{mp})
reads, 
\begin{equation}
-\mathop{\mathrm{Im}}\sigma=\varkappa_{dip}=%
-\frac{2 c V_{L}^{2}\varkappa_{dip}\ln\varkappa_{dip}}%
{1+4 V_{L}^{2}\varkappa_{dip}^{2}\ln^{2}\varkappa_{dip}},
\end{equation}
while the effective shift due to impurities is given by
\begin{equation}
\epsilon_{dip}(c)\equiv\mathop{\mathrm{Re}}\sigma=\frac{c V_{L}}%
{1+4 V_{L}^{2}\varkappa_{dip}^{2}\ln^{2}\varkappa_{dip}}.\label{displ}
\end{equation}
As usual, the characteristic concentration for this point in the
spectrum should be determined by the relation
$|\mathop{\mathrm{Re}}\sigma|=|\mathop{\mathrm{Im}}\sigma|$. This
condition yields,
\begin{equation}
c_{dip}=-4\epsilon_{r}^{2}\ln|\epsilon_{r}|. \label{cdip}
\end{equation}
For $c\ll c_{dip}$, the magnitude of the LDOS  at this specific energy
is rapidly increasing with impurity concentration,
\begin{equation}
\rho_{loc}(\epsilon_{dip}(c))\approx(2/\pi)%
\{\exp[-1/(2cV_{L}^{2})]/(2cV_{L}^{2})\},
\end{equation}
and then is reaching its maximum value at $c=c_{dip}$,
\begin{equation}
\rho_{loc}(\epsilon_{dip}(c_{dip}))=(2\pi|V_{L}|)^{-1}.
\end{equation}
According to Eq.~(\ref{displ}), the dip position is gradually
displaced approximately by the amount $cV_{L}$ with increasing the
impurity concentration for $c\ll c_{dip}$, and, finally, this dip
totally disappears from the LDOS at the impurity cite at
$c\sim c_{dip}$. The described detachment of the LDOS curve from the
energy axis can be distinctly seen in the inset of the Fig.~\ref{f2}. 

\paragraph{Discussion}
In a system with the linear dispersion, the GF variation with the
intersite distance has the characteristic radius that is proportional
to $1/|\epsilon|$. Since in a 2D system the average distance between
impurities is proportional to $1/\sqrt{c}$, the anticipated magnitude
of the critical concentration of the spectrum rearrangement is sitting
around $\epsilon_{r}^{2}$. Two critical concentrations (\ref{cr}) and
(\ref{cdip}) that were argued for above are bracketing from both sides
this crude estimation. There is an intimate interdependence of this
LDOS rearrangement and the rearrangement of the whole electron
spectrum, which have been studied in \cite{skrloc}. It is not 
difficult to verify that the validity criterion introduced in
\cite{skrloc} is completely supporting the results derived here. 
Within the plain impurity model adopted, the well--defined resonance
state is possible only for a strong impurity perturbation. However,
the restrictions imposed on the resonance appearance for, say, the
double impurity should not be that severe \cite{balcond,dbl1,dbl2},
while the main physics of the LDOS rearrangement must remain
substantially the same. Analogous effect can be achieved by a
tuning of the impurity--host hopping parameter.  

\paragraph{Conclusion.}
In summary, we have demonstrated that the concept of the LDOS rearrangement
is fully justified.  The sharper and the closer to the Dirac point is
the resonance, the lower is the critical concentration of the LDOS
rearrangement.  The respective critical concentration should not be
exceeded for the resonance peak and the antiresonance dip to be
discernible in the shape of the LDOS at the impurity site. We presume
that the LDOS rearrangement is not specific to graphene or related
systems with the Dirac dispersion, and should occur in virtually any
system that manifests impurity states in the close vicinity of the van
Hove singularities in its spectrum.  

\paragraph{Acknowledgments.}
One of the authors (Yu.~V.~S.) is grateful to B.~L.~Altshuler,
V.~G.~Baryakhtar and L.~A.~Pastur for useful discussions. This work
was partially supported by the Program of Fundamental Research of NAS
of Ukraine.

\end{document}